\documentclass[twocolumn,aps,prd,epsf,floatfix,superscriptaddress]{revtex4}

\usepackage{exscale}                  
\usepackage[intlimits]{amsmath}       
\usepackage{amsfonts}
\usepackage{amssymb,amscd}
\usepackage[dvips]{epsfig}                   
\usepackage{array}
\usepackage[usenames]{color}
\usepackage{graphicx}
\usepackage{bm}

\usepackage{graphics}
\usepackage{graphicx}
\usepackage{dcolumn}

\newcommand{\ar}{\arrowvert}
\newcommand{\ra}{\rangle}
\newcommand{\la}{\langle}

\newcommand{\cd}{\! \cdot \!}
\newcommand{\be}{\begin{equation}}
\newcommand{\ee}{\end{equation}}
\newcommand{\bea}{\begin{eqnarray}}
\newcommand{\eea}{\end{eqnarray}}
\newcommand{\ba}{\begin{eqnarray}}
\newcommand{\ea}{\end{eqnarray}}

\newcommand{\Tr}{\textrm{Tr}}

\begin{document}

\title{
Two infrared Yang-Mills solutions in stochastic quantization \\
and in an effective action formalism
}

\author{Felipe J. Llanes-Estrada}
\affiliation{Deptartamento de F\'isica Te\'orica I, Universidad Complutense de Madrid, 28040 Madrid, Spain}
\author{Richard Williams}
\affiliation{Deptartamento de F\'isica Te\'orica I, Universidad Complutense de Madrid, 28040 Madrid, Spain}
\affiliation{Institut f\"ur Physik, Karl-Franzens--Universit\"at Graz, Universit\"atsplatz 5, 8010 Graz, Austria}

\begin{abstract}
Three decades of work on the quantum field equations of pure Yang-Mills theory have distilled two families of solutions in Landau gauge. Both coincide for high (Euclidean) momentum with known perturbation theory, and both predict an infrared suppressed transverse gluon propagator, but whereas the solution known as ``scaling''
features an infrared power law for the gluon and ghost propagators, the ``massive'' solution rather describes the gluon as a vector boson that features a finite Debye screening mass. \\
In this work we examine the gauge dependence of these solutions by adopting stochastic quantization. What we find, in four dimensions and in a rainbow approximation, is that stochastic quantization supports both solutions in Landau gauge but the scaling solution abruptly disappears when the parameter controlling the drift force is separated from zero (soft gauge-fixing), recovering only the perturbative propagators;
the massive solution seems to survive the extension outside Landau gauge.
These results are consistent with the scaling solution being related to the existence of a Gribov horizon, with the massive one being more general. \\
We also examine the effective action in Faddeev-Popov quantization that generates the rainbow and we find, for a bare vertex approximation, that the the massive-type solutions minimise the quantum effective action.
\end{abstract}

\maketitle
\section{Introduction}

The Dyson-Schwinger equations (DSEs) for the gluon propagator in Landau gauge
have been the object of intense study in recent years. Two solutions are
known in several truncation schemes~\cite{Szczepaniak:2001rg}, that we will call ``scaling'' and
``massive''. The main difference between them is that the
first~\cite{von Smekal:1997vx,von Smekal:1997is} features an infrared
power-law behavior at low Euclidean $k^2$ while the second produces a
propagator reminiscent of Yukawa-like theory~\cite{Mandula:1987rh,Cornwall:1981zr,Dudal:2007cw,Dudal:2008sp,Aguilar:2008xm,Boucaud:2008ky,Aguilar:2011yb}, with a (gauge-dependent) gluon mass. Note however that the gluon's Poincar\'e representation is that of a massless vector boson with two polarization states, and that any reference
here to `mass' should be interpreted in the sense of an inverse Debye 
screening length~\cite{Alkofer:2011va}. Note that such confusion can be avoided by referring to the massive solution by the synonymous term ``decoupling''~\cite{Fischer:2008uz}.

Lattice computations currently seem
consistent with the second, massive
solution~\cite{Maas:2006qw,Cucchieri:2009zt,Ilgenfritz:2010gu,Bogolubsky:2009qb,Oliveira:2010xc,Oliveira:2011zn},
but the scaling solution remains of theoretical interest because of its
many desirable theoretical properties (the infrared exponents can be
quite approximately determined for the entire tower of Green's functions
without truncating the system~\cite{Alkofer:2004it,Fischer:2006vf}, 
the Kugo-Ojima confinement scenario~\cite{Watson:2001yv} is realized, the non-perturbative BRST quartet~\cite{Alkofer:2011pe}.)

\begin{figure}
\includegraphics*[width=8cm]{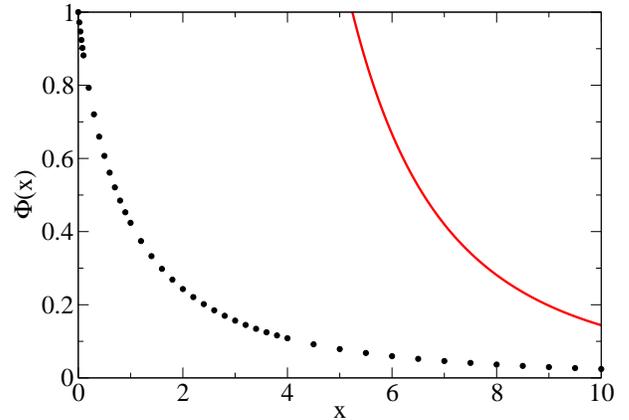}
\caption{\label{fig:ThomasFermi} Solid line (red online): Sommerfeld's
power-law solution of the Thomas-Fermi equation. Dotted line: numeric
solution to the equation imposing finiteness at the origin.}
\end{figure}

That these two types of solutions have been found for the Yang-Mills
system should not have come as a surprise, as the situation was similar
in atomic physics in the late 1920's. In the Thomas-Fermi~\cite{Thomas}
model of the atom, a central nucleus and a spherical electron cloud act
as sources of the electrostatic potential $V(r)$ (akin to the ghost
propagator in Coulomb gauge), $\Phi(x)$ in appropriately reduced units
\footnote{
For completeness we note that $x\equiv r/a$, with $a=a_0 (9\pi^2/(128Z))^{1/3}$, in terms of the nuclear charge $Z$ and Bohr's radius $a_0$,
and $\Phi\equiv rV(r)/(Ze)$ rescales the Coulomb potential due to the spherically symmetric (though not pointlike) charge distribution.
}
.
If the electron energy levels are filled according to the Pauli
principle in order of increasing kinetic energy alone, up to the Fermi
sphere, one obtains the Thomas-Fermi equation
\begin{align} \label{TFeq}
	\frac{d^2\Phi(x)}{dx^2} = \sqrt{\frac{\Phi(x)^3}{x}} \ .
\end{align}
If only one boundary condition to this equation is imposed for a neutral
atom, namely that the potential vanishes at large distance $r\propto x$,
one obtains the two well known solutions reproduced in
Fig.~\ref{fig:ThomasFermi}.
The solid line in the figure is Sommerfeld's solution,
$\Phi(x)=\frac{12^2}{x^3}$, which is the unique one with power-law
behavior. The dotted line represents the usually accepted numeric
solution modelling the atomic charge distribution, which is obtained
with the additional boundary condition that $\Phi(0)=1$ (other finite
boundary conditions lead to additional solutions that amount to a
rescaling of both $x$ and $\Phi$).  

Thus, it is natural that modern Yang-Mills Dyson-Schwinger equations or
Renormalization Group Equations, that are more sophisticated versions of
Eq.~(\ref{TFeq}), also accept solutions that are either finite or
power-law behaved in the infrared.

One of the questions we examine in this work is whether the existence of
the two solutions in Yang-Mills theory has any relation to the
well-known Gribov ambiguity: the Landau gauge fixing condition
$\partial_\mu A^\mu=0$ does not completely fix the gauge, but leaves a
discrete number of gauge ``Gribov'' copies~\cite{Zwanziger:1988jt}. Perhaps, one could
speculate, the two solutions correspond to different gauge
representations of the same gauge-invariant information, as it is known that the Gribov ambiguity causes a corresponding ambiguity in the DSE's~\cite{Zwanziger:2001kw,Maas:2009se,Maas:2011se}.

Thus, in the first part of this article we turn to a soft gauge-fixing method that avoids the Landau gauge condition and the Faddeev-Popov formalism, but produces Dyson-Schwinger equations that are quite similar in structure to the standard ones.

The basic observation of stochastic quantization~\cite{Parisi:1980ys} is
that the weight employed to compute correlators in Euclidean field
theory
\begin{align}
\la A(x) A(y) \ra = \int DA A(x) A(y) e^{-S_{\rm (YM)}[A]}\ ,
\end{align}
akin to a Boltzmann equilibrium distribution $e^{-\beta E}$,
can be seen as the end-point $e^{-S_{\rm (YM)}[A]}=\lim_{\tau\to\infty}
P(\tau)$ of the evolution of a stochastic random walk characterized by a
Fokker-Planck equation in a fictitious time parameter $\tau$. This is
\begin{align}
\label{fokplan}
\frac{\partial P}{\partial \tau} = \int d^4x \frac{\delta}{\delta A^{a \mu}(x)}
\left(
\frac{\delta P}{\delta A^a_\mu(x)} - K^a_\mu(x) P
\right) \ .
\end{align}
Formally, the force $K$ in this equation is defined to be $K^a_\mu(x)=-\frac{\delta S_{\rm (YM)}}{\delta A^a_\mu(x)}$.

However, Zwanziger showed that the limit $\tau\to\infty$ is not well
defined in a gauge theory because the stochastic evolution can run away
along a gauge orbit of gauge equivalent configurations $A(x)$. To avoid
it, one needs to add a force term that will pull the evolution towards
small values of $\int d^4x \ar A(x)\ar^2$ and thus retain the gauge
potential at (or near) the closest gauge copy to the origin in
$A$-space. This force has to be of dimension 3, to retain a hope of
renormalizability (the one-loop $\beta$ function has indeed been
calculated and found standard~\cite{MunozSudupe:1986ai,MunozSudupe:1986pb}, and the
Slavnov-Taylor identities obtained in~\cite{AlvarezEstrada:1988mz}), and
tangent to the gauge orbit to avoid disturbing any physical observables
(expectation values of gauge-invariant operators) that may be computed.

The natural choice is to add a gauge transformation
\begin{align}
K^a_\mu(x)=-\frac{\delta S_{\rm (YM)}}{\delta A^a_\mu(x)} + a^{-1} D^{ac}_{\mu}\partial\cd A^{c}(x) \ .
\end{align}
The parameter $a$ is a real and positive constant that controls the relative intensity of the stochastic Yang-Mills and the gauge-restoring forces. The gauge is not strictly fixed in stochastic quantization (rather, gauge-equivalent configurations will be weighted in a smooth manner around the origin in gauge space, with much less probability for those much farther) except in the limit $a\to 0$ that makes the gauge fixing force dominant and fixes the gauge field to Landau gauge.

The addition of this gauge force, tangent to the gauge orbit, makes the method of stochastic quantization geometrically correct, in the sense that the Gribov problem is bypassed.
If one then takes the limit $\tau\to \infty$ the path-integral weight $P$, gauge-equivalent to $e^{-S_{\rm (YM)}}$, is not easy to write down because the new force is not a conservative vector field in $A$-space, so it can not be written as a functional derivative of an action-like functional.
At least it has recently been shown that the Euclidean solution to the time-independent Fokker-Planck equation is positive and unique (up to normalization)~\cite{Vandersickel:2012tz}. In contrast, the solution in the Faddeev-Popov theory is highly non-unique owing to the nodes of the Faddeev-Popov determinant at each Gribov horizon~\cite{Zwanziger:2001kw}.

However one can undertake the computation of the quantum effective action and derive wave equations for the Green's functions of the theory.
Indeed the Dyson-Schwinger equations for the gluon propagator within the stochastic formalism have been obtained and described in~\cite{Zwanziger:2002ia}. 

A peculiarity of this ``soft'' gauge fixing method is that there are no
Faddeev-Popov ghosts. Instead one has both transverse and longitudinal
dressing functions of the propagator, with the relative weight $a$ being akin to
the gauge parameter.

The gluon propagator is
\begin{align}
\delta^{ab} D_{\mu\nu}(k)&= \int d^4 x \la A^a_\mu(0) A^b_\nu(x)\ra e^{ik\cd x} \, ,
\\ \label{deffullprop} 
D_{\mu\nu}(k) &= \frac{Z_T(k^2)}{k^2} P^T_{\mu\nu}(k)
 + a \frac{Z_L(k^2)}{k^2} P^L_{\mu\nu}(k) \, ,
\end{align}
with 
\begin{align}\label{eqn:projectors}
P^T_{\mu\nu} &\equiv  \left(\delta^{\mu\nu} -\frac{k^\mu k^\nu}{k^2} \right)
 \;\;\; , \;\;\;\;\;
P^L_{\mu\nu} \equiv  \frac{k^\mu k^\nu}{k^2} \ ,
\end{align}
appropriate transverse and longitudinal projectors.
We write down also the ``bare'' propagator from tree level perturbation theory, introducing appropriate renormalization constants for later use
\begin{align}
D^0_{\mu\nu}(k) = \frac{Z_3^{-1}}{k^2} P^T_{\mu\nu}(k)
+ a \frac{Z_a^{-1} Z_3^{-1}}{k^2} P^L_{\mu\nu}(k) \ .
\end{align}

We will recall the wave equations (in rainbow approximation to the DSE
of stochastic quantization) for $Z_L$ and $Z_T$ defined in
Eq.~(\ref{deffullprop}) in section ~\ref{DSEst}.  The DSE equations are
then a system of two coupled equations and we perform an infrared
analysis in section~\ref{IRanalysis}, in order to detect possible
infrared power-law solutions. This we perform in the limit of $a\to 0$
(Landau gauge) but also for finite $a$. In the first case we do find a
scaling solution, but not for finite $a$.

This is corroborated by our numerical solutions to the DSE system in section~\ref{NumSol}. We find the standard solution to the Landau gauge system in Faddeev-Popov quantization. Thereafter we solve the corresponding equations in Landau gauge in stochastic quantization. Finally we proceed beyond Landau gauge and find only the massive solution for finite $a$.

In section~\ref{FPaction} we return to the traditional method of Faddeev-Popov quantization, in Landau gauge, where both solutions are present. 
Our goal there is to construct an effective action for the traditional rainbow DSE's, and evaluate it numerically with the gluon and ghost propagator computed in section~\ref{NumSol}, with the idea of identifying the absolute minimum among them. 
We will find that the massive solution has least effective action, thus suggesting a reason why it should be found in lattice gauge theory simulations. It appears to us that a constrained minimization should be carried out in future work to find the scaling solution. What this constraint might be, is at present unknown to us, but our work on the stochastic quantization method suggests that it may be related to the formation of the Gribov horizon.

Finally, section~\ref{conclusions} presents a summary of our results and further discussion.

\section{Rainbow Dyson-Schwinger equation for $Z_L$, $Z_T$} \label{DSEst}

\begin{figure*}[!t]
	\begin{center}
\includegraphics*[width=0.95\columnwidth]{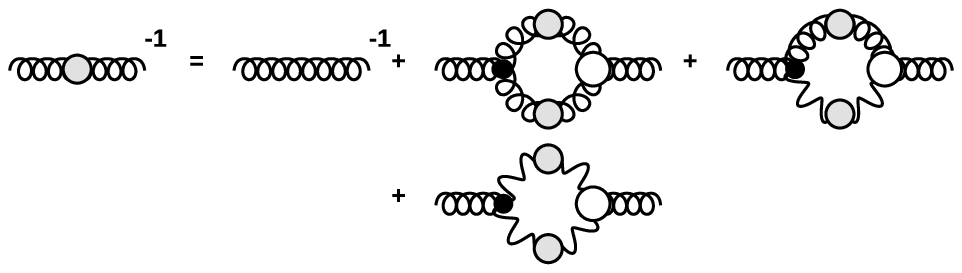}
\hfill
\includegraphics*[width=0.95\columnwidth]{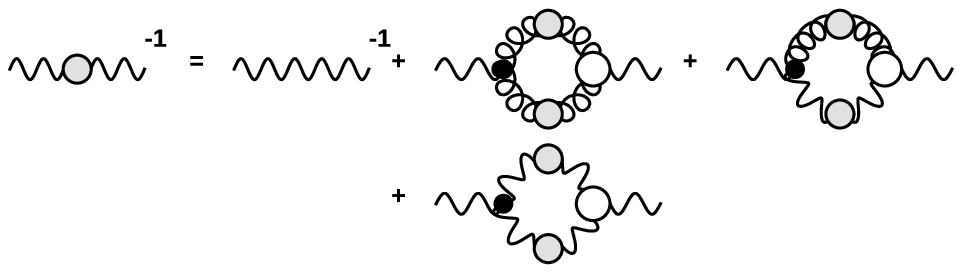}
\end{center}
\caption{\label{fig:picDSE} Diagrammatical representation of the DSE.
The presence of a transverse (longitudinal) projector is represented by a
spring (wave) in the adjoining gluons. The blobs indicate that the
propagator/vertex is dressed. We have subsumed numerical factors and
signs into the diagrams.}
\end{figure*}

The Dyson-Schwinger equation for the gluon propagator, after projecting
via the $P^T$ and $P^L$ of Eq.~(\ref{eqn:projectors}), becomes~\cite{Zwanziger:2002ia} a system of two
coupled equations, see Fig.~\ref{fig:picDSE}
\begin{align}
\frac{1}{Z_T(k^2)}    =& Z_3 - \frac{N_c}{6} Z_1 g^2  \nonumber     \\ 
&\times\int \frac{d^4k_1}{(2\pi)^4} 
\left( I^{TTT} + 2I^{TTL} + I^{TLL} \right)(k_1,k) \, , \nonumber\\[-2mm] \label{DSE1} \\[-2mm] \nonumber
\frac{1}{Z_L(k^2)} =& Z_3 Z_a - a Z_a  Z_1 g^2 \frac{N_c}{2} \nonumber \\
 &
\times\int \frac{d^4k_1}{(2\pi)^4} \left( I^{LTT}
+ 2I^{LTL} + I^{LLL} \right)(k_1,k) \, , \nonumber
\end{align}
\noindent where we employ slightly modified kernels $I$ respect
to~\cite{Zwanziger:2002ia} that we now specify; they do depend
non-linearly on $Z_L(k^2)$ and $Z_T(k^2)$ as usual.
To shorten the notation, we introduce $k_2=k-k_1$. Additionally a superscript
$T$ or $L$ above an index indicates that the index is contracted with
either of the projectors $P_T$ or $P_L$ with momentum of the
corresponding argument; the first of which corresponds to the external gluon with the second and third those gluons internal to the loop. For convenience we also write
$\tilde{Z}_3 = Z_3Z_a$ and $\tilde{a}= Z_a a$ from now on.
The six kernels read
\begin{align}
I^{TTT}(k_1,k) =& \frac{-Z_T(k_1^2) Z_T(k_2^2)}{k^2k_1^2k_2^2} \times \\ \nonumber
&  K^{TTT}_{\lambda\lambda_1\lambda_2}(k;-k_1,-k_2) G^{TTT}_{\lambda_1\lambda_2\lambda}(k_1,k_2,-k)
\\ 
I^{TTL}(k_1,k) =& \frac{-\tilde{a} Z_T(k_1^2) Z_L(k_2^2)}{k^2k_1^2k_2^2} \times \\ \nonumber
&  K^{TTL}_{\lambda\lambda_1\lambda_2}(k;-k_1,-k_2) G^{TLT}_{\lambda_1\lambda_2\lambda}(k_1,k_2,-k)
\\ 
I^{TLL}(k_1,k) =& \frac{-\tilde{a}^2 Z_L(k_1^2) Z_L(k_2^2)}{k^2k_1^2k_2^2} \times \\ \nonumber
&  K^{TLL}_{\lambda\lambda_1\lambda_2}(k;-k_1,-k_2) G^{LLT}_{\lambda_1\lambda_2\lambda}(k_1,k_2,-k)
\\
I^{LTT}(k_1,k) =& \frac{-Z_T(k_1^2) Z_T(k_2^2)}{k^2k_1^2k_2^2} \times \\ \nonumber
&  K^{LTT}_{\lambda\lambda_1\lambda_2}(k;-k_1,-k_2) G^{TTL}_{\lambda_1\lambda_2\lambda}(k_1,k_2,-k)
\\
I^{LTL}(k_1,k) =& \frac{-\tilde{a} Z_T(k_1^2) Z_L(k_2^2)}{k^2k_1^2k_2^2} \times \\ \nonumber
&  K^{LTL}_{\lambda\lambda_1\lambda_2}(k;-k_1,-k_2) G^{TLL}_{\lambda_1\lambda_2\lambda}(k_1,k_2,-k)
\\
I^{LLL}(k_1,k) =& \frac{-\tilde{a}^2 Z_L(k_1^2) Z_L(k_2^2)}{k^2k_1^2k_2^2} \times \\ \nonumber
&  K^{LLL}_{\lambda\lambda_1\lambda_2}(k;-k_1,-k_2) G^{LLL}_{\lambda_1\lambda_2\lambda}(k_1,k_2,-k) \ .
\end{align}
Before application of transverse and longitudinal projectors, the force
that generalises the bare vertex in the ordinary DSE's is
\begin{align}\label{BareForce}
	K_{\mu_1 \mu_2 \mu_3}(k_1;k_2,k_3) &= - S^{\rm (YM)}_{\mu_1\mu_2\mu_3}(k_1,k_2,k_3) \\
	& + \tilde{a}^{-1} K^{\rm (GF)}_{\mu_1\mu_2\mu_3}(k_1;k_2,k_3) \,,\nonumber
\end{align}
with
\begin{align} \label{3gtree}
	-S^{\rm (YM)}_{\mu_1\mu_2\mu_3}(k_1,k_2,k_3) &= \left[ (k_1)_{[\mu_2}\delta_{\mu_3]\mu_1}+{\rm
	(cyclic)} \right]\, , \\ \label{3gtree2}
	K^{\rm (GF)}_{\mu_1\mu_2\mu_3}(k_1;k_2,k_3)&= \left[\left((k_3)_{\mu_3}\delta_{\mu_1\mu_2} -(2\leftrightarrow 3)
	\right) \right] \, .
\end{align}
These contain the standard three-gluon vertex and the Gauge-Fixing term. 
While the action part is Bose symmetric, the gauge force distinguishes
its first index. Further, note
that in contrast to Ref.~\cite{Zwanziger:2002ia} we have pulled out the
common factor of $(ig)$ to the front of the DSE equation.\\
An immediate remark regarding Eq.~(\ref{3gtree2}) is that
$K^{LTT}=K^{TTT}=0$ in light of $K$ being proportional to either $k_2$ or
$k_3$ that have zero projection onto the transverse plane.

Next we specify the dressed vertex $G$, that essentially defines the
truncation of the DSE system (we work only at one loop in the elementary
vacuum polarization insertion).
The $G$ vertex is a sum of a bare three-gluon vertex 
Eq.~(\ref{3gtree}) (that would generate the rainbow approximation to the pure Yang-Mills system) and a term appropriate to stochastic quantization that incorporates the $a$-dependent drift force,
\begin{align}\label{TotalForce}
	G_{\mu_1\mu_2\mu_3}(k_1,k_2,k_3) &= S^{\rm (YM)}_{\mu_1\mu_2\mu_3}(k_1,k_2,k_3)\nonumber\\
&+\Gamma_{\mu_1\mu_2\mu_3}(k_1,k_2,k_3)\ .
\end{align}
where once again, in contrast to Ref.~\cite{Zwanziger:2002ia} we have
pulled out the common factor $(ig)$ to the front of the DSE.

We project each of the indices by transverse/longitudinal projectors.
These quantities are anti-symmetric in their three arguments and so
$\Gamma^{TTL}_{\mu_1\mu_2\mu_3}(k_1,k_2,k_3) =
-\Gamma^{TLT}_{\mu_1\mu_3\mu_2}(k_1,k_3,k_2)$, and likewise
$\Gamma^{LLT}_{\mu_1\mu_2\mu_3}(k_1,k_2,k_3) =
-\Gamma^{TLL}_{\mu_3\mu_2\mu_1}(k_3,k_2,k_1)$. These stochastic vertices
stem from truncated solutions of the quantum effective action and are
found to be
\begin{align}
\Gamma^{TTT}_{\mu_1\mu_2\mu_3}(k_1,k_2,k_3) =& 0 \, ,\\ 
\Gamma^{TTL}_{\mu_1\mu_2\mu_3}(k_1,k_2,k_3) =&  \frac{k_2^2-k_1^2}{\tilde{a} (k_1^2+k_2^2) + k_3^2 } (k_3)_{\mu_3} \\ \nonumber 
 \times&\left[P^T(k_1) P^T(k_2)\right]_{\mu_1 \mu_2} \,,    \\ \label{GammaTLL}
\Gamma^{TLL}_{\mu_1\mu_2\mu_3}(k_1,k_2,k_3)=&\frac{-1}{\tilde{a}} \bigg( \frac{k_3^2-\tilde{a}k_1^2}{\tilde{a}k_1^2+k_2^2+k_3^2}(k_2)_{\mu_2} \\ \nonumber
\times&\left[P^T(k_1)P^L(k_3)\right]_{\mu_1\mu_3} - (2\leftrightarrow
3)\bigg)\,, \\
\Gamma^{LLL}_{\mu_1\mu_2\mu_3}(k_1,k_2,k_3)=& \frac{-1}{\tilde{a}} \bigg(
\frac{k_2^2-k_3^2}{k_1^2+k_2^2+k_3^2}(k_1)_{\mu_1}\\ \nonumber
\times& \left[P^L(k_2)P^L(k_3)\right]_{\mu_2\mu_3} +({\rm cyclic})\bigg)
\,.
\end{align}

The conventional DSE system of equations equivalent to Eq.~(\ref{DSE1}) in
the Faddeev-Popov formalism in Landau gauge is long known to have two
solutions~\cite{Boucaud:2008ji,Boucaud:2008ky}. The massive
solution has a transverse Yukawa-like gluon propagator $Z(k^2)/k^2 \to
c/(k^2+m^2)$, and the scaling solution instead takes a power-law form
$Z(k^2)/k^2 \to (k^2)^{2\kappa-1}$, that, for the (slightly truncation dependent) $\kappa\simeq 0.595$ of Faddeev-Popov theory, yields a very suppressed gluon propagator, while
the ghost propagator is enhanced $G(k^2)/k^2\to (k^2)^{-\kappa-1}$.

A practical way~\cite{Fischer:2008uz} to numerically select one or the other solution is to subtract the second equation of Eq.~(\ref{DSE1}) at a fixed scale $k=0$ to facilitate imposing a boundary condition, obtaining 
\begin{align} \label{DSE2}
\frac{1}{Z_L(k^2)} &=  \frac{1}{Z_L(0)} + \\ \nonumber
&\tilde{a} Z_1 g^2 \frac{N_c}{2} \int \frac{d^4k_1}{(2\pi)^4} 
\left( I_l(k_1,k)-I_l(k_1,0)\right) \, ,
\end{align}
(where for brevity we have subsumed the three kernels in the integrand
into one function $I_l$). Note that subtracting at zero is not
necessary, rather it increases stability of the numerical solution.

One can now choose  $\frac{1}{Z_L(0)}$ to be finite (in search for a
massive-like solution) or zero (in search of a scaling solution with
divergent ghost dressing function), and try to see whether the system
does accept both solutions, in analogy with the Thomas-Fermi equation,
and now in the absence of the Gribov ambiguity.

\section{Infrared Analysis} \label{IRanalysis}
To examine the structure of the integral equations near $a=0$, in search for an extension of the Landau gauge power-law scaling solution for the propagators, it is useful to
classify the various terms according to their power of $a$.  
Recall that the vertices $K$ and $G$, in Eqs.~(\ref{BareForce}) and (\ref{TotalForce}) respectively,
contain a Yang-Mills part $S^{(\rm{YM})}$ together with a gauge transformation part ($K$ or $\Gamma$) that can
introduce a non-trivial $a$ dependence. For brevity, we write the
kernel product
$K^{TTT}_{\lambda\lambda_1\lambda_2}(k,-k_1,-k_2)G^{TTT}_{\lambda_1\lambda_2\lambda}(k_1,k_2,-k)$
as $K^{TTT}_{\lambda\lambda_1\lambda_2}G^{TTT}_{\lambda_1\lambda_2\lambda}$
where the ordering dictates the function arguments. Then for the
transverse gluon we have
\begin{align}\label{ITTTa}
I^{T,TT}(k_1,k) =&\frac{Z_T(k_1^2) Z_T(k_2^2)}{k^2k_1^2 k_2^2}   S_{{\rm (YM)}}^{TTT} S_{{\rm (YM)}}^{TTT}
	    \,\, ,
\end{align}
since $K^{{\rm (GF)}}_{TTT} = 0$ and $\Gamma^{TTT}=0$.
For $I^{T,TL}$ we find
\begin{align}\label{ITTLa}
I^{TTL}(k_1,k) =& -\frac{ Z_T(k_1^2) Z_L(k_2^2)}{k^2k_1^2k_2^2} \times \\ \nonumber
& \bigg[ \tilde{a} \left(-S_{{\rm (YM)}}^{TTL}S_{{\rm (YM)}}^{TLT}
-S_{{\rm (YM)}}^{TTL}  \Gamma^{TLT}\right) \\ \nonumber
& + \left(  K_{{\rm (GF)}}^{TTL}S_{{\rm (YM)}}^{TLT}
+K_{{\rm (GF)}}^{TTL}  \Gamma^{TLT}\right) \, ,
\end{align}
and finally,
\begin{align}\label{ITLLa}
I^{TLL}(k_1,k) =& -\frac{ Z_L(k_1^2) Z_L(k_2^2)}{k^2k_1^2k_2^2} \times \\ \nonumber
& \bigg[ \left( K_{{\rm (GF)}}^{TLL}  \widetilde{\Gamma}^{LLT}\right) + \tilde{a}^2 \left(-S_{{\rm (YM)}}^{TLL}S_{{\rm (YM)}}^{LLT}
\right) \\ \nonumber
& + \tilde{a} \left( -S_{{\rm (YM)}}^{TLL} \widetilde{\Gamma}^{LLT} +K_{{\rm
(GF)}}^{TLL}S_{{\rm (YM)}}^{LLT} \right)\bigg] \,\,.
\end{align}
In this last equation we have defined $\widetilde{\Gamma}^{LLT}=\tilde{a}\Gamma^{LLT}$ in order to factor out the explicit factor
$\tilde{a}$ in Eq.~(\ref{GammaTLL}).

For the longitudinal gluon propagator, we also include the overall factor of
$\tilde{a}$ present in Eq.~(\ref{DSE1}) to make the counting more
transparent.  We have
\begin{align}\label{ILTTa}
	\tilde{a} I^{LTT}(k_1,k) =& -\frac{ Z_T(k_1^2) Z_T(k_2^2)}{k^2k_1^2k_2^2} \times \\ \nonumber
	& \bigg[ \tilde{a}\left( -S_{{\rm (YM)}}^{LTT} S_{{\rm
	(YM)}}^{TTL}-S_{{\rm (YM)}}^{LTT} \Gamma^{TTL} \right) \bigg] \,\,,
\end{align}
having used $K_{{\rm (GF)}}^{LTT} = 0$ to simplify the expression.

For $I^{LTL}$ we write
\begin{align}\label{ILTLa}
	\tilde{a} I^{LTL}(k_1,k) =& -\frac{ Z_T(k_1^2) Z_L(k_2^2)}{k^2k_1^2k_2^2} \times \\ \nonumber
	& \bigg[ \left(  K_{{\rm
	(GF)}}^{LTL}\widetilde{\Gamma}^{TLL}\right) + \tilde{a}^2 \left( -S_{{\rm (YM)}}^{LTL} S_{{\rm
	(YM)}}^{TLL}\right) \\ \nonumber
	& + \tilde{a} \left( K_{{\rm (GF)}}^{LTL} S_{{\rm (YM)}}^{TLL} -
	S_{{\rm (YM)}}^{LTL} \widetilde{\Gamma}^{TLL} \right)\,\, ,
\end{align}
once again writing $\Gamma^{TLL}=\tilde{a}\widetilde{\Gamma}^{TLL}$.

Finally,
\begin{align}\label{ILLLa}
	\tilde{a} I^{LLL}(k_1,k) =& -\frac{ Z_L(k_1^2) Z_L(k_2^2)}{k^2k_1^2k_2^2} \times \\ \nonumber
	& \bigg[ \tilde{a}\left( K_{{\rm (GF)}}^{LLL}
	\widetilde{\Gamma}^{LLL}\right)
	+ \tilde{a}^3 \left( -S_{{\rm (YM)}}^{LLL} S_{{\rm
	(YM)}}^{LLL}\right) \\ \nonumber
	& + \tilde{a}^2 \left( K_{{\rm (GF)}}^{LLL} S_{{\rm (YM)}}^{LLL} -
	S_{{\rm (YM)}}^{LLL} \widetilde{\Gamma}^{LLL} \right) \bigg]
	\,\,,
\end{align}
In this last expression we also accounted for factors of $\tilde{a}$ in
the vertex by writing $\Gamma^{LLL}=\tilde{a}\widetilde{\Gamma}^{LLL}$.

It is now a simple matter to read off terms that survive in the Landau
gauge limit, which we investigate in the next section.

\subsection{Landau Gauge}

%
After substituting the kernels arranged in powers of the parameter $a$ (that quantifies the separation from Landau gauge) in Eqs.~(\ref{ILTTa})--({\ref{ILLLa}) into the Dyson-Schwinger equations (\ref{DSE1}), and
reading off the surviving terms in the $\tilde{a}\rightarrow 0$ limit,
we find for the longitudinal gluon
\begin{align}
	\frac{1}{Z_L(k^2)} =& \frac{1}{Z_L(0)} - \frac{g^2N_c}{\left( 2\pi
	\right)^4k^2} \int d^4k_1 \\ \nonumber
	& \times\Bigg(\frac{2 \left[ k^2k_1^2-\left( k\cdot k_1
	\right)^2 \right]}{k_1^4\left( k_2^2+k^2 \right)}Z_T(k_1^2)
	Z_L(k_2^2)\Bigg)\,\,,
\end{align}
that has already been reported in Ref.~\cite{Zwanziger:2002ia} where
similarity with the ghost DSE of Faddeev-Popov gauge has been noted.

If a scaling solution should exist, the infrared behavior of the dressing functions would respectively be 
\begin{align} \label{powerlaws}
	Z_T(k^2) \sim&\, C_T (k^2)^{\alpha_T}\,\,,  \nonumber \\[-2mm]
	\\[-2mm] \nonumber
	Z_L(k^2) \sim&\, C_L (k^2)^{\alpha_L}\,\, . 
\end{align}
Substituting into the longitudinal gluon equation~(\ref{DSE1}) and employing dimensional analysis for small $k$,
with the integration dominated by the region of small $k_1\propto k$, we find 
$\alpha_T +2\alpha_L = -(4-d)/2$ for $d$-spacetime dimensions, or in four dimensions simply
the traditional Landau gauge result from Faddeev-Popov quantization
$$
\alpha_T=-2\alpha_L\ .
$$

For the transverse gluon we must consider the kernels of
Eqs.~(\ref{ITTTa})--({\ref{ITLLa}) in Eq.~(\ref{DSE1}) upon taking the
$\tilde{a}\rightarrow 0$ limit. Two of these contributions are given in
Ref.~\cite{Zwanziger:2002ia}
\begin{align}
	I^{T,TT}&(k_1,k) =
	-\frac{1}{k^2}\frac{Z_T(k_1^2)}{k_1^2}\frac{Z_T(k_2^2)}{k_2^2}\times \\
	&S^{  {\rm(YM)} TTT}_{\lambda\lambda_1\lambda_2}(k,-k_1,-k_2)
	S^{ {\rm(YM)} TTT}_{\lambda_1\lambda_2\lambda}(k_1,k_2,-k)
\,\,,
\end{align}
and
\begin{align}
	I^{T,LL}(k_1,k) = -2 \frac{k_1^2k^2 - \left( k_1\cdot k
	\right)^2}{k^4 k_1^2 k_2^2}Z_L(k_1^2)Z_L(k_2^2)\,\,.
\end{align}
These two integral kernels are identical to the gluon- and ghost-loop
kernels in standard Faddeev Popov theory in the bare-vertex truncation.
However, in Landau gauge one contribution that was previously overlooked
stems from the mixed longitudinal/transverse gluon correction
\begin{align}
	I^{T,TL} =& 2\,K^{TTL}_{\lambda\lambda_1\lambda2}(k;-k_1,-k_2)
	S^{TLT}_{\lambda_1\lambda_2\lambda}(k_1,k_2,-k) \\
	&\nonumber\times
	\frac{1}{k^2}\frac{Z_T(k_1^2)}{k_1^2}\frac{Z_L(k_2^2)}{k_2^2}
	\,\,.
\end{align}
This arises from the identification $\Gamma^{TLT}_{\lambda_1\lambda_2\lambda}(k_1,k_2,-k)) =
S^{TLT}_{\lambda_1\lambda_2\lambda}(k_1,k_2,-k))$ in the limit
$\tilde{a}\rightarrow0$, and hence contributes additively in
$G^{TLT}=S^{TLT}+\Gamma^{TLT}=2S^{TLT}$ (this corrects an error in Ref.~\cite{Zwanziger:2002ia}, where these terms were taken to cancel, which however did not affect the calculation reported there).

Performing the $k_1$ integrals analytically assuming they are dominated by the infrared power laws in Eq.~(\ref{powerlaws}), we find
$\alpha_L \equiv -\kappa \simeq -0.52146$.
This is consistent with an IR enhanced longitudinal gluon propagator
$Z_L/k^2 \propto (k^2)^{-1.52}$ and an IR suppressed transverse gluon propagator $Z_T/k^2 \propto (k^2)^{0.04}$, similar to the Faddeev-Popov ghost-gluon system in Landau gauge but with a slightly smaller exponent.

\subsection{IR analysis for $a\ne0$}
We now attempt to find a scaling solution for finite $a$.
To focus only on the important terms, we assume infrared dominance of the gauge-fixing ``force'' $K^{gt}_\mu(A)$ over the Yang-Mills ``force'' $-\delta S_{\rm YM}/\delta A$, in analog to the Faddeev-Popov ghost dominance in Ref.~\cite{Alkofer:2004it}. Thus we analyse the effect on the integral equation of the gauge-fixing force alone,
\begin{equation}
	K_\mu(A) =\tilde{a}^{-1} K_{gt,\mu}\ .
\end{equation}
With this restriction, triply transverse vertices $K^{T,TT}$ and
$\Gamma^{T,TT}$ vanish, as does $K^{L,TT}$. The surviving terms in the
DSEs are
\begin{align}
	a^{-1} \frac{1}{Z_L} =& I^{L,LL} + I^{L,TL} \\
	\frac{1}{Z_T} =& I^{T,LL} + I^{T,TL}
\end{align}
Substituting again the power law ans\"atze in equation~(\ref{powerlaws})
(with the assumption that no leading-power cancellation occurs), we find
\begin{align}
	\alpha_T = \alpha_L \, ,
\end{align}
since independently of which term on the RHS may dominate we find the
same power on the RHS of both equations for $L$ and $T$. By counting
powers of $k^2$
\begin{align}
-\alpha_L = d/2 - 2(1+\alpha_L) \;,
\end{align}
which yields $\alpha_L=(4-d)/6$ for Euclidean dimension $d$. Thus
\begin{align}
	\alpha_T=&\alpha_L=0\qquad       &\textrm{for}\, d=4 \nonumber &\ ,\\
	\alpha_T=&\alpha_L=1/6           &\textrm{for}\, d=3 \nonumber &\ ,\\
	\alpha_T=&\alpha_L=1/3           &\textrm{for}\, d=2 \nonumber &\ .
\end{align}
Thus, in four dimensions we find that the only power-law solution is consistent with the perturbative propagators proportional to $1/k^2$, the dressing not carrying an anomalous infrared dimension\footnote{As a side remark we comment that the disappearance of the scaling
solution in four dimensions in the setting of stochastic quantization is
perhaps expected because of the analysis of~\cite{von Smekal:1997vx} since
the scaling solution seems to be a feature of ghost-antighost symmetric
Faddeev-Popov quantization, and such symmetry is not implemented in the
stochastic scheme.}.

Other studies concerning scaling behaviour of the Yang-Mills system in gauges different to Landau gauge have been performed,  such as ghost-antighost symmetric gauges~\cite{Alkofer:2003jr}, maximally Abelian gauges~\cite{Huber:2009wh,Alkofer:2011di} and interpolating gauges~\cite{Fischer:2005qe}.  Similarly, linear covariant gauges have also been investigated on the lattice~\cite{Cucchieri:2008zx,Cucchieri:2011pp}. Typically these introduce an additional scale which dominates in the IR. However, note that the naive scaling analysis does not preclude the existence of a mixed scaling regime, in which the scaling behaviour is only manifest at intermediate, rather than infrared, momenta. An example of this scenario is QED in 2+1 dimensions~\cite{Fischer:2004nq}.

Upon numerically solving the complete system of equations, see Eq.~(\ref{DSE1}), matching the infrared and the ultraviolet behavior with a computer, we expect to find both the massive and scaling solutions for $a=0$, but perhaps only the massive solution for $a\ne 0$. We thus turn to the computer in the next section.

\section{Numerical Solution of the DSE equations in stochastic quantization}
\label{NumSol}

As a warm-up we first show the solution to the conventional system of the coupled one-loop gluon-ghost DSE equations that has been discussed at length elsewhere recently~\cite{Fischer:2008uz}. Note here that in contrast to recent studies, we employ the bare vertex approximation since we will later investigate the effective action for this Faddeev-Popov system. In contrast to Ref.~\cite{Atkinson:1997tu} which also employed the bare vertices approximation, we will not employ an angular approximation.

For the sake of clarity, we briefly overview the solution method. For
the scaling solution, one assumes the existence of power-law behaviour
for the ghost and gluon dressing functions:
\begin{equation}
	G(p^2) = A p^{-\kappa}\,\,\,\, Z(p^2) = B p^{2\kappa} \, ,
	\label{eqn:power}
\end{equation}
with the precise value of $\kappa$ dependent upon the assumption of
ghost dominance and the Taylor condition $\widetilde{Z}_1=1$. Through
the running coupling associated with the ghost-gluon vertex one surmises the
existence of an infrared fixed point in the coupling. This gives a
stringent relationship between the coefficients $A$ and $B$ in 
Eq.~(\ref{eqn:power}). Typically, one chooses a value for $g^2$ and $A$,
with $B$ then determined through knowledge of the IR fixed point. 
The renormalization constants for the ghost
and gluon propagators are removed by subtraction,
in favour of renormalization conditions for $Z(\mu^2)$ and $G(\mu^2)$.

The ghost equation must be subtracted at $\mu^2 = 0$ zero 
to choose the boundary condition, as in Eq.(\ref{DSE2}). To search for the scaling solution we impose the condition $1/G(0)=0$ in accordance with Kugo-Ojima. 
The final condition for $Z(s^2)$ is then surmised by smooth matching of the numerical
solutions to the IR. Ultimately,
$\mu$ is then arbitrary and one scales the solutions in the momentum
variable to match say the physical value of the running coupling at
some scale. 

The resulting ghost and gluon dressing function for both massive and
scaling solutions are represented in Fig.~\ref{fig:conventionalDSE}.

\begin{figure*}[!t]
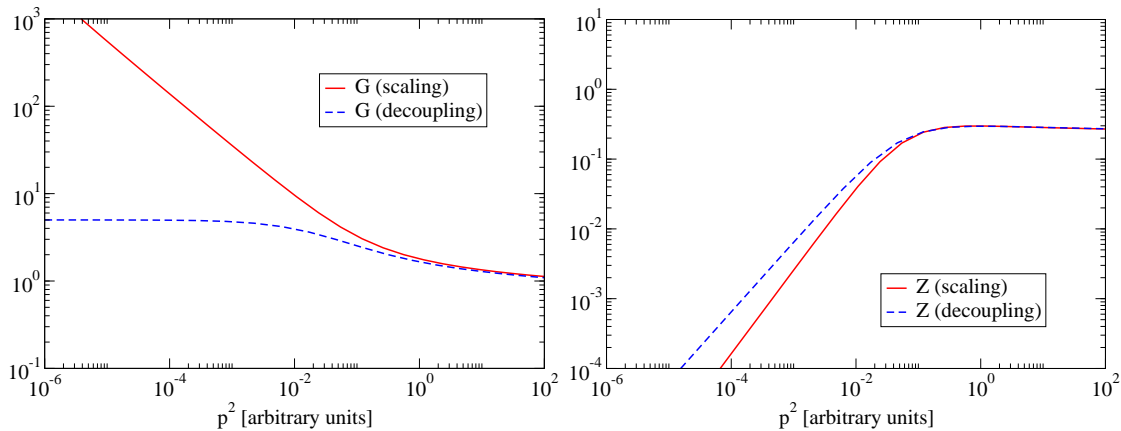

\includegraphics*[width=0.85\columnwidth]{ym_ghost}
\includegraphics*[width=0.85\columnwidth]{ym_gluon}
\caption{\label{fig:conventionalDSE} Numerical solution to the system of
coupled Faddeev-Popov ghost-gluon equations at one loop (rainbow
approximation with bare vertices, in Landau gauge). Both scaling and
massive solutions are represented. Left graph: ghost dressing function.
Right graph: gluon propagator dressing function.}
\end{figure*}

In the case of Stochastic QCD we follow the procedure as closely as
possible. The longitudinal gluon is subtracted at zero momenta, whilst
the transverse gluon is subtracted at some large perturbative scale.
In the Landau gauge limit one similarly has access to an IR fixed point.
By employing bare-vertices we lose Multiplicative Renormalizability and
so, it is less trivial to relate the choice of the 
renormalization condition $Z(s)$ to the momentum scale. 

To address the massive solutions, one merely replaces the boundary
condition for $G(0)=\textrm{finite}$. Note that we could perform the
subtraction of the ghost and gluon equations at the same large
perturbative momentum, obtaining both massive and scaling solutions.
In this case fine-tuning of the renormalization condition dictates which type of solution
is selected.

One technicality of this particular system of equations is the
appearance of quadratic divergences in the transverse equation for the
gluon propagator. These are eliminated by merely employing an additional
subtraction in the infrared. 

In Fig.~\ref{fig:decoupling} we then plot the analogous quantities in
the framework of stochastic quantization in Landau gauge, the
longitudinal and transverse dressing functions, setting $a\to 0$ (Landau
gauge).
\begin{figure*}[!t]
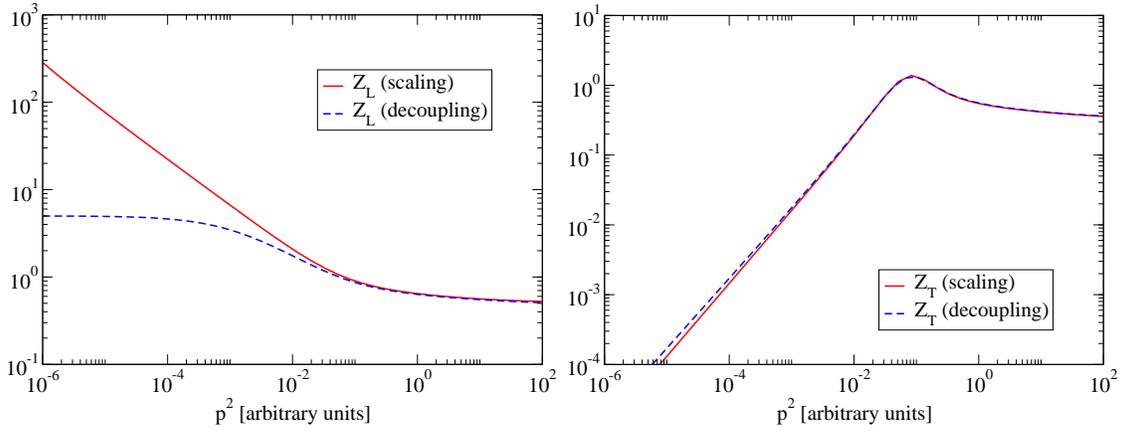

\centering
\includegraphics*[width=0.85\columnwidth]{gluonL}
\includegraphics*[width=0.85\columnwidth]{gluonT}
\caption{\label{fig:decoupling} Dressing function of the longitudinal  
gluon (left-panel) and transverse gluon (right-panel) for both the scaling- and decoupling-type solutions, in stochastic quantization with $a=0$ (Landau gauge).}
\end{figure*}
The numerical result for the scaling solution fulfills our expectations
based on the analytical infrared scaling study, with the correct
$\kappa$ value that we obtained analytically. The massive solution seems
a simple modification of that obtained in Faddeev-Popov quantization,
which comes as no surprise as the two systems of equations in Landau
gauge, whether stochastic or Faddeev-Popov, are very similar.

Finally we proceed to finite (but very small) $a$ in Fig.~\ref{fig:stochasticDSE}.
\begin{figure*}[!t]
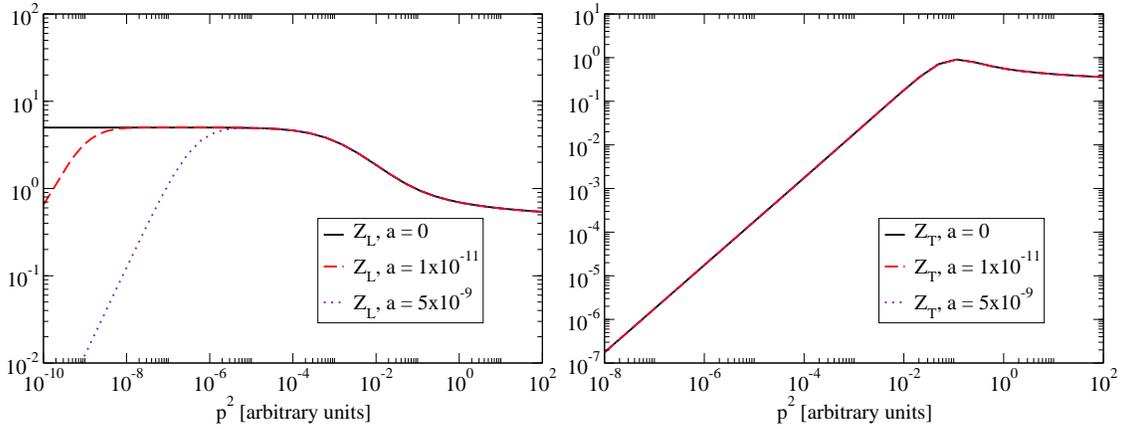

\includegraphics*[width=0.85\columnwidth]{gluonLa}
\includegraphics*[width=0.85\columnwidth]{gluonTa}
\caption{\label{fig:stochasticDSE}Numerical solution to the system of
coupled longitudinal-transverse gluon equations at one loop (rainbow
approximation with bare vertices, in stochastic quantization). The
scaling solution now has disappeared (only the perturbative propagator
comes out of the analysis), while the massive solution still exists.
Left graph: longitudinal dressing function $Z_L$. Right graph:
transverse gluon dressing function $Z_T$.}
\end{figure*}

Our numeric results confirm that the  scaling solution with a suppressed transverse gluon is a feature of stochastic QCD only in Landau gauge, and that a massive-type
solution can however be found even with soft gauge fixing. This is in agreement with extant lattice data~\cite{Pawlowski:2009iv,Aiso:1997au}.

\section{Effective Action for the Faddeev-Popov system} 
\label{FPaction}
In this section we return to Faddeev-Popov quantization of Yang-Mills theory and
consider it well established that functional methods find two solutions to
the wave equations. We plot these in a simple truncation in
Fig.~\ref{fig:conventionalDSE}.

It is thus puzzling that lattice gauge theory reproduces only one of them. 
A natural avenue of investigation is to consider an effective quantum
action from which the functional (be it DSE or Renormalization Group
equations) are derived, and study its value for the different solutions.
Perhaps lattice gauge theory is picking up the absolute minimum of the
effective action and the second, scaling solution, can only be found
with a constrained minimization. Recently, a study of the renormalization group~\cite{Weber:2012vf} singled out the `massive' solutions as those stable in the infrared.  Here, we wish to investigate whether any of these solutions is a local minimum of the effective action, or
rather if they are all saddle points.

A convenient starting point is the two-particle irreducible (2PI)
quantum effective action of Yang-Mills theory to one-loop. Propagators
in the effective action are the ones in the fully interacting vacuum,
but the vertices are taken from perturbation theory. Thus, taking a
functional derivative respect to the explicit propagators produces the
rainbow Dyson-Schwinger equations to one loop without need to worry
about implicit dependences of the vertices. It is for this reason 
that we employed the bare-vertex approximation in the study above. 
The effective action reads~\cite{Berges:2004pu}
\begin{align} \label{effaction}
	\Gamma\left( D,G \right) &= \frac{i}{2}\Tr \log D^{-1} +
	\frac{i}{2}\Tr\left( D_0^{-1}D \right) \nonumber\\
	&-i\Tr\log G^{-1} -i\Tr\left( G_0^{-1}G \right) + \Gamma_2[D,G] \,
	,\nonumber\\
\end{align}
where the non-trace part involving higher than 2-point functions is given as
\begin{align} \label{effg2}
	\Gamma_2[D,G] = +\frac{i}{12}g^2D^3V_{03}^2 -
	\frac{i}{2}g^2DG^2V_{03}^{(gh)2} \,,
\end{align}
and represented diagrammatically in Fig.~\ref{fig:efaction}.
\begin{figure}
	\begin{align*}
		\Gamma_{2} = 
		\frac{i}{12}
		\begin{array}{c}
			\includegraphics[scale=0.8]{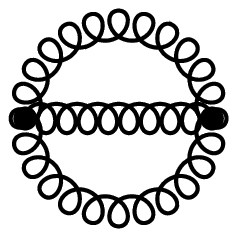}
		\end{array}
		-\frac{i}{2}
		\begin{array}{c}
		\includegraphics[scale=0.8]{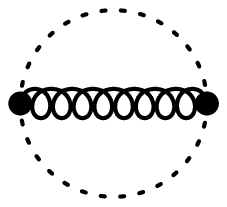}
		\end{array}
	\end{align*}
\caption{Non-trace part (involving dressed three-point function) of the effective action that generates the rainbow Dyson-Schwinger equations of Yang-Mills theory by seeking the stationary point under variation of $G$, $Z$ (diagrammatically, cutting through each line).\label{fig:efaction}}
\end{figure}
The vertices here are $V_{03}$, that is the three-gluon vertex in perturbation theory (at this order identical with the Lagrangian-level vertex in Eq.~(\ref{3gtree}) $S^{YM}_{\mu_1\mu_2\mu_3}({\bf k}_1,{\bf k}_2,{\bf k}_3)$) and the bare ghost-gluon vertex $V_{03}^{(gh)}$.

The full gluon and ghost propagators in Landau gauge are symbolically represented by $D$ and $G$, and their perturbative counterparts by $D_0$ and $G_0$.

If one takes a functional derivative of Eq.~(\ref{effaction})
respect to the propagators $\delta \Gamma/ \delta D=0$, $\delta \Gamma/ \delta G=0$, the rainbow one-loop Dyson-Schwinger equations for $D$ and $G$ follow.
For example, let us derive the abstract expression for the ghost equation,
\begin{align}
	\frac{\delta\Gamma(D,G)}{\delta G} &= \nonumber -i G_0^{-1} -\frac{i}{G^{-1}}\left( -G^{-2}
	\right)-\frac{i}{2}g^2D\cdot
	2GV_{03}^{(gh)2}  \nonumber \\
	&= 0 \ .
\end{align}
This gives
\begin{equation}
	G^{-1}= G_0^{-1} + g^2 D G V_{03}^{(gh)2} \, ,
\end{equation}
and likewise, for the gluon equation
\begin{equation}
	D^{-1}= D_0^{-1} + \frac{g^2}{2} D^2 V_{03}^{2}\ .
\end{equation}

With the propagators parametrized in terms of the conventional gluon and ghost dressing functions
\begin{align}
D_{\mu\nu}(k) &= \frac{Z(k^2)P^T_{\mu\nu}(k^2)}{k^2} \, ,\\
G(k) &= \frac{G(k^2)}{k^2}\ ,
\end{align}
the effective action is a functional of $Z(k^2)$ and $G(k^2)$ that has zero variation at the solution of the Dyson-Schwinger equations.
The reduction of the effective action to a simple form that can be
handled by a computer given $Z$, $G$ is shown in
appendix~\ref{app:action}. The outcome is, up to constant contributions,
\begin{align}
\Gamma =& 8 V \int\frac{d^4k}{(2\pi)^4} \bigg\{ \log
\left(G^{-1}(k^2)\right)
+G(k^2)   \\ \nonumber 
&\qquad\qquad-\frac{3}{2} \left[\log
\left(Z^{-1}(k^2)\right)+Z(k^2) \right] \\ \nonumber
&- \frac{3}{2}g^2\int\frac{d^4q}{(2\pi)^4} K_1(k,q) G(k^2) G(q^2) Z((k-q)^2)
\\ \nonumber 
&-g^2\int\frac{d^4q}{(2\pi)^4} K_2(k,q) Z(k^2) Z(q^2)Z((k-q)^2)
\bigg\} \, ,
\end{align}
with kernels
\begin{align}
	K_1(k,q)\equiv& \frac{k^2q^2-k \cdot q^2}{k^2 q^2 r^4}  \\
	K_2(k,q)\equiv& \frac{k \cdot q^2-k^2q^2 }{k^4q^4r^4}
	\left[ 3r^2\left(k^2+q^2\right)+2k^2q^2 + k\cdot q^2 \right] ,\nonumber
\end{align}
where $r = k-q$. 

A very appealing way of visualizing the effective action is to take a
curve in function space that passes by the two solutions of the DSE
equations in Landau gauge (massive and scaling). 
The $OY$-axis of Fig.~\ref{fig:effaction} is the effective action
$\Gamma_\Lambda(D,G)$ for cutoff $\Lambda$, whereas the $OX$ axis
represents an arbitrary interpolation parameter $\alpha$ that varies
between $0$ and $1$.

So that we can show also the propagators on the perturbative vacuum
state $G_0$, $Z_0$, we choose a parabola through all three functions
(we actually employ an interpolation of their logarithm since the functions are very dissimilar, and all three are positive)
\begin{align} 
	\log \widetilde{Z}[\alpha] =& \left( 2\alpha-1 \right)\left[ \left( \alpha-1
	\right)\log Z_0+\alpha \log Z_2 \right] \nonumber \\ 
                               &-4\alpha\left( \alpha-1 \right)    \log Z_1\, , \\
	\log \widetilde{G}[\alpha] =&\left( 2\alpha-1 \right)\left[ \left( \alpha-1
	\right)\log G_0+\alpha \log G_2 \right] \nonumber\\ 
                               &-4\alpha\left( \alpha-1 \right)\log G_1 \,,
\end{align}
obtaining, presumably a local maximum at $Z_0$ and then an absolute
minimum at either of $(Z_1,G_1)$ or $(Z_2,G_2)$.

We thus proceed to express the effective action $\Gamma$ in terms of
generic ghost and gluon dressing functions, $Z$ and $G$, with the eye on
evaluating it over the parabola passing by the three relevant choices.
We will drop all constant contributions to the action, as we are
interested in the relative ordering of the solutions and not the value
of the action itself, which is of little relevance.

\begin{figure}
\includegraphics[width=0.95\columnwidth]{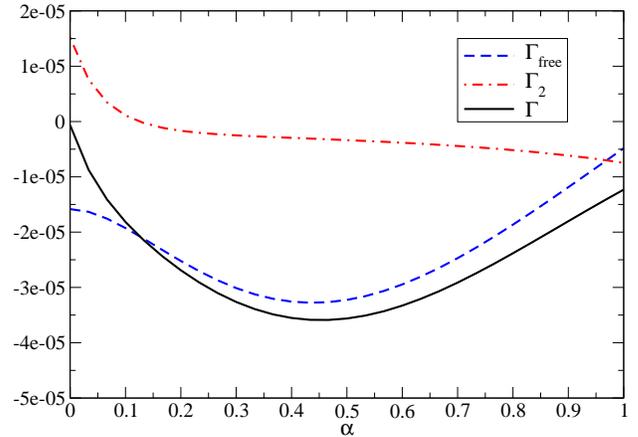}
\caption{\label{fig:effaction} 
The solid line depicts the 2PI cut-off quantum effective 
action (units $GeV^4$ per unit four-volume)
that generates rainbow-DSE equations, plotted on a slice through the
function space of ghost and gluon Euclidean momentum propagators. For
$\alpha=0$ the propagator is bare, for $\alpha=1/2$ its dressing
functions yield the massive solution, and for $\alpha=1$ we arrive to
the scaling solution. We choose a cut-off that emphasises the difference
in the infrared. }
\end{figure}

Our numeric results are shown in Fig.~\ref{fig:effaction}.
 
We separately show the contribution from the kinetic terms ($\Gamma_{\rm free}$, denoted with a dashed line) that features a maximum of the action (along this slice) at the perturbative propagator (on the left of the plot at $\alpha=0$). This is natural as this function satisfies the free Dyson-Schwinger equation, thus it must be an extremum of the free effective action. 

That is no more the case for the full computation including interactions (solid line) where the perturbative propagator plays no special role. However a clear minimum can be seen very near the massive solution at $\alpha=1/2$, that seems to be the extremum of the interacting action.  The scaling solution is not an extremum of the unconstrained minimization. Note that one should not attribute meaning to the precise location of the minimum in Fig.~\ref{fig:effaction} as this depends upon the interpolating function used. It is merely indicative of which class of solutions minimises the action.

Finally, the figure separately depicts $\Gamma_2$, the interacting part of the effective action at two loops, as defined in Eq.~(\ref{effg2}), that is the difference between the total and the free actions.
In conclusion, from an analysis of the effective action in bare-vertex approximation, it appears that the massive solution to the Dyson-Schwinger equations 
is naturally found in lattice gauge theory due to its lesser action. 
Whether the scaling solution can be found by an adequately constrained minimization, as well as a more extensive exploration of the effective action, deserves another publication.

\section{Summary} \label{conclusions}

We have addressed the contemporary topic of the two solutions in the Dyson-Schwinger equations of pure Yang-Mills theory, after pointing out that finding both finite and infrared-scaling solution has really been with us since the time of Thomas' analysis of the atom, since Sommerfeld actually found a power-law solution to the Thomas-Fermi equation for the electrostatic potential there in addition to the well-known finite solution.

In Yang-Mills theory which of these two solutions is found depends on the chosen gauge. We have employed Landau gauge to show within a standard rainbow
truncation scheme, in conventional Faddeev-Popov quantization, that the massive solution has less effective action than the scaling solution, and thus we conjecture that this is the reason why the massive (or decoupling) solution to the gluon DSE is so easily found in lattice gauge theory. Whether this remains to be the case for the dressed system is yet to be seen, but we expect the qualitative picture to remain the same.

For our main contribution we have adopted the viewpoint of stochastic quantization, where a gauge fixing is not forced, but instead the system adopts a statistical distribution where, in each gauge trajectory, it tries to relax to the gauge copy closest to the origin. The control parameter of this thermodynamic gauge force, $a$, when set to 0, allows to guarantee the Landau gauge condition in stochastic quantization. Within this Landau gauge, we find results similar to the Faddeev-Popov method, with both scaling and massive solutions.

If this condition is lifted, so that Landau gauge is not fixed, by allowing $a$ to be finite, the scaling solution to the Dyson-Schwinger equation abruptly
disappears. The massive family of solutions remains as the only one for finite $a$. Thus we have another piece of evidence that might suggest that the scaling solution is related to the Gribov horizon forming in the curvilinear Landau gauge.

In conclusion, we believe, from this and other works, that there are indeed two classes of solutions to the complete wave equations of Yang-Mills theory, as found in various truncations of the Dyson-Schwinger and the Exact Renormalization-Group equations. 
We also conjecture that lattice gauge theory is finding the solution with least effective action in an unconstrained minimization, which is a massive-like solution, and that the scaling solution probably needs to be obtained with a minimization that takes into account additional constraints.  
Our study of the two solutions with the stochastic quantization method finds that the scaling solution is a property of Landau-gauge fixing, and disappears if configurations not in Landau gauge are allowed to contribute to the path integral. This gauge dependence is in line with other findings in the literature~\cite{Maas:2008ri}. We conjecture here that the scaling solution is related to the formation of the Gribov horizon and that perhaps this is the direction to search for an appropriate boundary condition that restricts the lattice configurations in order to also find a possible scaling solution in Monte Carlo simulations of Yang-Mills theory.

\begin{acknowledgments}
We wish to thank D.~Zwanziger for his continued support and contributions to this work. We also thank R.~Alkofer, C.~Fischer and L.~von Smekal for discussions and a critical reading of this manuscript. This work has been supported by grants FPA2011-27853-01, FIS2008-01323 (Spain) and by the Austrian Science Fund FWF under Project M1333-N16.
\end{acknowledgments}

\appendix
\section{Evaluation of the Effective Action for Faddeev-Popov quantization in Landau gauge}\label{app:action}

\subsection{Evaluation of the free action}
We first deal in this subsection with the terms that are of zero order in the strong coupling $g$.

Consider the two terms of Eq.~(\ref{effaction})
that include the logarithm of an inverse propagator $D_{\mu\nu}$ or $G$. Note that $\log{A}$ is best interpreted in the diagonal basis
\begin{equation}
	\log\left[\begin{array}{ccc}
		A_1 &       &     \\
		    &\ddots &     \\
		    &       & A_n \\
		     \end{array}\right]
	=
	\left[\begin{array}{ccc}
		\log{A_1} &       &     \\
		    &\ddots &     \\
		    &       & \log{A_n} \\
		     \end{array}\right]\, ,
\end{equation}
so that, for  a constant times the identity, $\log\left( c \delta^{ab}
\right)=\left(\log{c}\right)\cdot \delta^{ab}$. Likewise, $\log\left( f(a)\delta^{ab}
\right)=\left(\log(f(a))\right)\delta^{ab}$. 

We can easily calculate the traced logarithm of the bare ghost propagator
$G_0^{-1} = G_0^{-1\,ab}(x,y) = i\Box_x\delta(x-y)\delta^{ab}$ to be
\begin{equation}
	-i\Tr\log{G_0^{-1}} 
	=-i\sum_a \int d^4x \delta^{aa}\left( \log{i\Box_x}
	\right)\delta(x-y)\bigg|_{x=y} \ .
\end{equation}
The constant contribution $\log(ix) = \log{i} + \log{x}\equiv \log{x}$
can be neglected, and we represent the delta function by 
\begin{equation}
	\delta(x-y)\equiv \int \frac{d^4k}{\left( 2\pi
\right)^4}\exp\left[-ik(x-y)\right]\ ,
	\label{eqn:deltaexp}
\end{equation}
that takes an additional factor $i$ in Euclidean space because $d^4k\to id^4k_E$.
Since $\sum_a\delta^{aa}=8$, $\int d^4x =
V$ with $V$ the space-time volume we find
\begin{equation}
	-i\Tr\log{G_0^{-1}} =  8 \cdot V \int \frac{d^4k}{\left(
	2\pi \right)^4}\log(k^2) + \textrm{constant}
\end{equation}

Now we dress the ghost propagator, $k^2 \rightarrow
\frac{k^2}{G(k^2)}$  (denoted in co-ordinate space by a
tilde, $\Box_x\rightarrow \widetilde{\Box}_x$) and obtain 
\begin{equation} \label{FirstCont}
	-i\Tr\log{G^{-1}} = 8V \int \frac{d^4k}{\left( 2\pi
	\right)^4}\log\frac{k^2}{G(k^2)} \, .
\end{equation}

Next we address the traced logarithm of the gluon propagator,
$\frac{i}{2}\Tr \log{D^{-1}}$. Starting again with its bare counterpart
and ignoring constant contributions,
\begin{align}\nonumber
	&\frac{i}{2}\Tr \log{D_0^{-1}} \!=\! 
        \frac{i}{2}\Tr\log\left[ (-i)\!\left(
	\delta_{\mu\nu}\Box-\partial_\mu\partial_\nu \right)_x
	\delta^{ab}\delta(x-y)
	\right]
	\\ 
        &\!\to 4i \!\sum_{\mu=1}^4 \delta^{\mu\nu}\!\!\!\int\!\! d^4x\log\left[ \left(
	\delta_{\mu\nu}\Box-\partial_\mu\partial_\nu \right)_x
	\right]\delta(x-y) \bigg|_{y=x} \ .
\end{align}
Once again we replace the delta function by Eq.~(\ref{eqn:deltaexp})
and set $x=y$ to leave
\begin{align}
\frac{i}{2}\Tr \log{D_0^{-1}} &= \\ \nonumber
	\frac{-8 }{2}&\!\!\int\! d^4x \sum_{\mu=1}^4
	\delta^{\mu\nu}\!\!\int\!\frac{d^4k}{(2\pi)^4}
	\log\left[ -\delta_{\mu\nu}k^2 +k_\mu k_\nu \right]\, .
\end{align}
We write $\log\left[ -\delta_{\mu\nu}k^2 +k_\mu k_\nu
\right]=\log(-1)+\log(k^2\hat{\delta}_{\mu\nu}(k))$ in
terms of the projector transverse to $k$,
$\hat{\delta}_{\mu\nu}(k)=\delta_{\mu\nu}-\hat{k}_\mu\hat{k}_\nu=\log(k^2)\hat{\delta}_{\mu\nu}(k)$
(it is not difficult to convince oneself that this relation is valid in
the $\perp_k$ subspace writing the expression in coordinates. The
constant $\log(-1)$ is dropped). Thus
\begin{align}\nonumber
\frac{i}{2}\Tr \log{D_0^{-1}} &=\frac{-8}{2} \cdot V \int d^4k \underbrace{\sum_{\mu=1}^4
	\hat{\delta}_{\mu\nu}(k)}_{3} \log{k^2} \\
	&= \frac{-24}{2}  \cdot V \int \frac{d^4k}{\left( 2\pi
	\right)^4}\log{k^2} \, .
\end{align}
Proceeding to the dressed gluon similar to that of the ghost yields,
\begin{equation}
	\frac{i}{2}\Tr\log{D^{-1}} = -12 V \int
	\frac{d^4k}{\left( 2\pi \right)^4}\log\left( \frac{k^2}{Z(k^2)} \right)\ .
\end{equation}

Next we address the quadratic terms in the free action,
\begin{align}
       -i&\Tr \left( G_0^{-1}G)\right) =  \nonumber\\
	 &-i \sum_{ab}\int\!\!\int d^4x d^4y G_0^{-1\,ab}(x,y)G^{ba}(y,x) =\nonumber\\
	\nonumber& -8i \int\!\!\int d^4x d^4y 
	\frac{\Box_x}{\widetilde{\Box}_y}\int\!\!\int \frac{d^4k}{\left( 2\pi
	\right)^4} \frac{d^4q}{\left( 2\pi \right)^4}\cdot
	e^{-i(k-q)(x-y)}\,. \\  
\end{align}
Note the `dressed' d'Alembertian to represent the inverse propagator in
position space. Because of translational invariance we make a change of
variables with unit Jacobian, $z=x-y$, $z'=\frac{x+y}{2} $, with the
integral over $z'$ yielding the integration volume $V$. 
Thus after simplification with the Euclidean conventions we have
\begin{equation}
	i\Tr\left( G_0^{-1}G \right)= 8 V \int
	\frac{d^4k}{\left( 2\pi \right)^4}G(k^2) \, .
\end{equation}

Likewise we construct
\begin{align}
  \frac{i}{2}&\Tr\left( D_0^{-1}D \right) = \nonumber \\ \nonumber
	&\frac{i}{2}\sum_{ab}\sum_{\mu\nu}\int\!\!\int d^4x d^4y \bigg( \delta^{ab}\left( \delta_{\mu\nu}\Box-\partial_\mu\partial_\nu
        \right)_x\delta(x-y)
\\ 
	&\qquad\qquad\times\delta^{ba} 
       \left(\widetilde{\delta_{\nu\mu}\Box-
                        }\partial_\nu\partial_\mu
	\right)^{-1}_y
\delta(x-y) \bigg)\,.
\end{align}
Passing to Fourier momentum space and simplifying
\begin{align}
	\frac{i}{2}\Tr\left( D_0^{-1}D \right)= \frac{-8}{2}V\int \frac{d^4k}{\left( 2\pi \right)^4}\left(
	1+3\cdot Z(k^2) \right)\,.
\end{align}

Collecting the free parts of the effective action together we find
\begin{align}
	\Gamma^{\rm free} = 8V\int\!\!\frac{d^4k}{\left( 2\pi
	\right)^4}&\bigg\{ 
	\log\left( \frac{k^2}{G(k^2)} \right)+G(k^2) \\ \nonumber  &  
	\!\!\!\!\!\!\!\!-\frac{3}{2}\log\left( \frac{k^2}{Z(k^2)}
	\right)-\frac{1+3Z(k^2)}{2}
	\bigg\} \ ,
\end{align}
in which we can discard the constant contribution to the effective
action that results from $\log(k^2)$.

It is easy to check that this free effective action, upon minimization,
demands the $G$ and $Z$ ghost and gluon dressing functions to take
their tree-level value
\begin{align}
	0=\frac{\delta\Gamma^{free}}{\delta G(q^2)} &= const. \left(
	-G_q^{-1}+1
	\right) \\ \nonumber &\Rightarrow \;\;\;\;G_{free}(q^2)=1\, , \\
	0=\frac{\delta\Gamma^{free}}{\delta Z(q^2)} &= const. \left(
	\frac{3}{2}Z_q^{-1}-\frac{3}{2}
	\right) \\ \nonumber &\Rightarrow \;\;\;\;Z_{free}(q^2)=1\,.
\end{align}
\subsection{$g^2$ part of the effective action}
We now turn to the interacting part of the action, Fig.~\ref{fig:efaction}.
In rainbow approximation it is given by the sum of two terms,
$\frac{i}{12}g^2D^3V_{03}^2$ and $-\frac{i}{2}g^2DG^2 V_{03}^{(gh)2}$.
Let us start by computing the ghost loop
\begin{align}
	-&\frac{i}{2}g^2DG^2 V_{03}^{(gh)2} =  \\ \nonumber
        &-\frac{i}{2} g^2
\left[ \int i \frac{d^4k}{\left( 2\pi \right)^4}
	\right]
	\left[ \int i \frac{d^4q}{\left( 2\pi \right)^4}\right]
	\left( i f^{abc}k^\mu \right)\left( i f^{acb} q^\nu \right)
\\ \nonumber 
	&\!\!\!\!\frac{iG(k^2)}{k^2} \frac{i G(q^2)}{q^2} \frac{i Z( \left( k-q
	\right)^2}{\left( k-q \right)^2}\left( \delta^{\mu\nu} -
	\frac{(k-q)^\mu(k-q)^\nu}{\left( k-q \right)^2} \right)\,,
\end{align}
where we employ the conventions of Ref.~\cite{Berges:2004pu}.
Summing Euclidean and color indices, 
\begin{align}
	-\frac{i}{2}g^2DG^2 V_{03}^{(gh)2} &= \\ \nonumber
	&\!\!\!\!\!\!12 g^2 \int 
	\frac{d^4k}{\left( 	2\pi \right)^4}
	\frac{d^4q}{\left( 2\pi \right)^4}
	\frac{ k^2 q^2 - \left( k\cdot q \right)^2 }{k^2 q^2\left( k-q \right)^4} \\ \nonumber
&\times G\left( k^2 \right) G\left( q^2	\right) Z\left(  \left( k-q \right)^2 \right)
\ .
\end{align}
Because of invariance under rotations, the external $k$-integral can be reduced to $2\pi^3 \int_0^\infty dk k^3$, and the remaining invariance under two rotations of the internal $q$-integral
(whose integrand depends on $\cos\theta = \widehat{k\cdot q}$ since $k$ has already been fixed) simplifies to
$4\pi \int_0^\infty dq q^3\int_0^\pi d\theta \sin^2\theta$
which leaves a total of three integrations to eventually be performed on a computer.

Whereas the momentum integrals display a quadratic divergence that need to be regulated consistently with the DSE, the angular integral is convergent at end points due to the $\sin^4\theta$ factor.

We reduce the gluon loop following the same procedure.

After simplifying (with the aid of FORM~\cite{Kuipers:2012rf}), 
\begin{align}
 \frac{i}{12}&g^2 D^3 V_{03}^2= \frac{8 g^2}{\left( 2\pi
 \right)^8}\int\!\!\int d^4k d^4q
	\frac{Z(k^2)Z(q^2)Z(r^2)}{k^4 q^4 r^4}
\\ \nonumber
	\times&\bigg\{
\left(k \cdot q^2-k^2q^2\right)
	\left( 3r^2\left(k^2+q^2\right)+2k^2q^2 + \left(k\cdot q\right)^2 \right)
	\bigg\}\,,
\end{align}
where $r = k-q$
Again, this is an eight-dimensional integral featuring quadratic
divergences in each k, q; it is reducible to 3D as before.

\end{document}